\documentclass[aps,prb,twocolumn,groupedaddress,floatfix]{revtex4}
\usepackage{graphicx}
\usepackage{dcolumn}
\usepackage{bm}

\begin{document}

\title{Effect of the equilibrium pair separation on cluster structures}

\author{Y. Yang and D. Y. Sun}
\affiliation{State Key Laboratory of Precision Spectroscopy and
Department of Physics, East China Normal University, Shanghai
200062, China}

\date{\today}

\begin{abstract}

A simple pair potential, which equilibrium pair separation can be
varied under a fixed interaction range, has been proposed. The new
potential can make both face-centered-cubic(fcc) and body-centered-cubic(bcc)
structure stable by simply changing one parameter. To investigate
the general effect of the potential shape on cluster structures, the
evolution of cluster structures is calculated for different
equilibrium pair separations. The small size clusters($N<25$), which
adopt the polytetrahedra, are almost independent on the details of
the potential. For the large size clusters($25<N<150$), the
potential with large equilibrium pair separation trends to stable
decahedra and close-packed structure, disordered clusters appear for
the potential with small equilibrium pair separation, while for the
middle range of equilibrium pair separation, the icosahedra are
dominated.

\end{abstract}

\pacs{}

\maketitle

\section{Introduction}

Nanoclusters, with the exotic physical and chemical properties, have
brought a big chance for future industry in many areas, it also
challenges our current theories and models for traditional condensed
matters.\cite{Nalwa04} Until now, many phenomena of nano-system
still puzzle physicists, chemists and material scientists.
Undoubtedly, any precise description of cluster properties needs the
correct structural models. To date few direct measurements of cluster
structures are available experimentally,\cite{Marks94_exp_Dh} and
much of the current theoretical understanding of cluster structures
has been derived from atomic-scale molecular-dynamics(MD) and Monte
Carlo(MC) simulations.\cite{Baletto05_review}

In computer simulations, finding the most stable structure
corresponds to search the global minimum of complicated
multi-dimensional potential energy surfaces(PES).\cite{Wales03_PES} The
potential energy surfaces of clusters are usually represented by an
appropriate energetic model, such as that based on \emph{ab-initio}
method, tight-binding(TB) models, or (Semi)empirical interaction
potentials. The model clusters, described by Lennard-Jones(LJ) and
Morse potentials, have been studied extensively(e.g.,
Refs.\cite{Baletto05_review} and references therein). Metal clusters
are also widely studied using both DFT and classical many-body
potentials(e.g., Refs.\cite{Baletto05_review,Longo05_Mna,Longo05_Mnb} and references
therein). The studies were even extended to large molecule
systems\cite{Doye95b_c60,Doye96_c60,Doye01_c60,
Hernandez04_c60ion,Hernandez06_c60water,Hodges00_water,Maheshwary01_abinit_water,
Wales98_water,Lee01_water,James05_water,Gonzalez05_waterion} and
multi-matter
system.\cite{Jellinek96_NiAl,Rossi04_AgNi,Sabo04_XY,Rapallo05_AgX,Doye05a_blj,
Ferrando05_bimetal,Longo05_NiFe,Longo06_FeMn,Curley07_AgAu}
Many exotic structures, which is probably forbidden in bulk
materials, have been reported in the previous studies. The excellent
example includes, the cage structures of
carbon,\cite{Kroto85_fulleren} the
Icosahedra(\emph{IH})\cite{Mackay62_ih,Martin96_ih} and
decahedra(\emph{DH})\cite{Ino69_Dh,Marks84_Dh} based atomic shell
structures of metal clusters, and the recent found cage structures
of gold clusters,\cite{Gu04_dft_au_cage,Ji05_dft_au_cage}\emph{etc}.

Indubitably the cluster structures are determined by the details of
the interatomic interaction. One of the current focuss is to examine
the general structural effects of the different contributions to the
interaction. Doye and Wales found that the Friedel type oscillation
in atomic potentials can strongly modulate the cluster
structures.\cite{Doye01a_dzugutov,Doye01b_dzugutov,Doye03_dzugutov}
The original and mechanism of disordered structure in metal clusters
were studied in term of the different many-body
forms.\cite{Soler00,Doye03_Gupta} Baletto \emph{et al} have shown
how the potential forms affect the crossover size between different
structural motifs (icosahedra, decahedra, and truncated
octahedra).\cite{Baletto02_wulf_eam_ag} Michaelian \emph{et al} have
made a comparison of the global minimum structure for different
many-body potentials.\cite{Michaelian99_eam_au}
 Doye and Wales investigated the structural
consequence issue for a set of Sutton-Chen families of
potentials.\cite{Doye98_S-C} Gong and his coworkers have studied the
relativistic effect on the structure of gold clusters, which leads
the presences of cage-like
structures.\cite{Gu04_dft_au_cage,Ji05_dft_au_cage}

For studying the general effects of potential shapes on cluster
structures, Doye and Wales suggested to consider a potential which
is simple enough that one can comprehend the effects of any changes
to its form. This method has been used to investigate the effect of
the potential range and anisotropy on the cluster structures. Braier
\emph{et al} first made the study on six- and seven-atom clusters
bound by the Morse potential for different interaction
range.\cite{Braier90_morse} A similar study was preformed on other
model potentials. Rey and Gallego have shown how the structures and
melting behavior of hard-core Yukawa clusters changing with the
range of attractive Yakawa tail.\cite{Ray96_Yukawa_range_pot} Based
on the generalized LJ, Morse, and Born-Mayer potentials, Amano
\emph{et al} recently investigated the structure change with the
potential shape.\cite{Amano06_range} Doye and Wales made a
systematic search of the PES of Morse clusters as a function of the
interaction range. They found that, decreasing the range results in
destabilizing strained
structures.\cite{Doye95_morse,Doye96a_morse,Doye96b_morse,Doye97_morse}
The general trend have been used to explain the growth sequences of
other systems.\cite{Doye95b_c60}

Despite substantial efforts by many researchers, our knowledge about
the relationship between cluster structure and potential shape is
still limited on a few factors, specially the effect of the
interaction range.
Although the interaction range does play an important role for an
interatomic potential, other factors also should not be neglected.
Obviously the further studies along this line are needed. Another
important factors could be the equilibrium pair
separation($d_{EPS}$), which reflect the size of atoms. Specially,
for a fixed interaction range, how the equilibrium pair
separation($d_{EPS}$) affect the cluster structure is worthy to
answered. Physically, changing of $d_{EPS}$ corresponds to changing the
atomic size. The importance of the atomic size issue has been shown
in recent studies on $C_{60}$.\cite{Hagen93_c60} In present paper,
by introducing a simple model potential, which has the fixed
interaction range but variable equilibrium pair separation, or
equally speaking 'the atomic size', we have studied the effect of
$d_{EPS}$ on the cluster structures.

The remainder of the paper is organized as follows. The following
section describes the new developed model potential and
computational details. The cluster structures for a few selected
parameters are presented in Sec. III. The conclusions drawn from
this work are summarized in Sec. IV.

\section{Computational details}

The potential we used was originally obtained from the effective
pair potential\cite{Carlsson90_effective} of EAM potential for
iron,\cite{Ackland97_FeEAM,Mendelev03_FeEAM} then we re-parameterize it
in the current form:

 \begin{eqnarray}
  \Phi(r_{ij})&=\left\{\begin{array}{ll}
                \varepsilon [\frac{\displaystyle r_{ij}}{\displaystyle \sigma}-2.5]^{3}[\gamma-1.44719(\frac{\displaystyle \sigma}{\displaystyle r_{ij}})];r_{ij}\leq2.5\sigma  \\
                0;r_{ij}>2.5\sigma  \\
                \end{array} \right.
 \end{eqnarray}

where $\varepsilon$ and $\sigma$ is energy and length unit
respectively. $r_{ij}$ denotes the distance between atom \emph{i}
and \emph{j}. $\gamma$ is an adjustable parameter, which determines
the $d_{EPS}$ of the potential. The higher value of $\gamma$ gives a
shorter $d_{EPS}$. $\varepsilon$ is chosen to keep the potential
well depth equal to one for each $\gamma$, and $\sigma$ equals to
one for all case studied. Figure 1 shows the new potential with a
few selected $\gamma$ of 0.8,0.9,0.95,1.0,1.05,1.1,1,2. For
comparing, this figure also shows the Morse and LJ potential, where
they have been fitted to have the same curvature at the bottom of
the potential well as the present one. From this figure, one can see
that, changing $\gamma$ corresponds to the alternation of the
equilibrium pair separation, also can be regarded as changing the
atomic size. Of course, by changing $\gamma$, the stiffness of the
potential is also changed, which is similar to the Morse potential
in this point. Comparing to Morse and LJ potential, the new one is
softer in repulsive part and stiffer in attractive part, and shorter
in interaction range.

Another feature of the current potential is that it can make both
fcc and bcc stable by varying $\gamma$. Table 1 presents the
cohesive energy of fcc and bcc phase as a function of $\gamma$. The
both fcc and bcc have the same energy at zero temperature at
$\gamma$=1. For $\gamma$ larger than one, bcc phase is stable over
fcc, while the smaller $\gamma$ favors fcc phase. The reason is
that, for large $\gamma$, the potential well becomes more and more
flat. In this case, the dominated interaction in bcc is the both
first and second nearest neighbors totally 14 atoms, while only 12
first nearest neighbors are contributed to energy in fcc due to the
limited interaction range. It needs to point that the Morse
potential always make the fcc more stable over bcc. This is the major
difference between the present potential and Morse one. This potential
is also similar to the Johnson potential for bcc
Iron,\cite{Johnson89_potential} which implies that this form may
represent some major physics of bcc based metals.

To optimize the cluster structure, we first search the global
minimum among all the known structures for each potential and
cluster size\cite{CambridgeClusterDatabase} using the steepest-descent
method. Then we make further exploration for most stable structure
with the generalized-simulated-annealing algorithm, which has
been shown in previous work as a powerful and efficient
procedure.\cite{XiangSFG97_gsa} The most stable structure then can
be found among these structures.

\section{Results and discussions}

Fig.2 plots the second difference of
energy($\Delta_{2}E=E(N+1)+E(N-1)-2E(N)$) as a function of cluster
sizes for all studied $\gamma$. From top to bottom, it corresponds
to the $\gamma$ from 0.8 to 1.2 respectively. For size smaller than
24, all the curves have the same trend. This implies that the
structures are less dependent on the details of potentials at small
sizes, which is similar to the case in Morse
potentials.\cite{Doye95_morse} For all $\gamma$ expect 0.8, around
size from 135 to 147, $\Delta_{2}E$ is almost equal to zero. This is
because these clusters are based on $\emph{IH}_{147}$ cluster by taking
a few atoms among its 12 vertex atoms. These vertex atoms are all
equal due to the symmetry. As the size larger than 24, the
differences of these curves begin to emerge. The curve of
$\gamma$=0.8 is evidently different from others for size larger than
23. $\gamma$=0.9 and 0.95 are similar each other in whole range of
sizes. The major trends of $\gamma$=1.0, 1.05, 1.1 and 1.2 are close
in the most ranges. These differences are the reflection of
different structure type as discussing below.

Peaks in $\Delta_{2}E$ correspond to clusters which are stable
compared to adjacent sizes and have been found to correlated with
magic numbers in mass spectra of clusters. We note that there are
two types of peaks in $\Delta_{2}E$. One type of peaks corresponds
to especially stable clusters, such as N=13,55,147, the closed-shell
\emph{IH} structures. Another one is directly related to the change
of structural types. For example, $\Delta_{2}E$ for $\gamma$=0.8 has
several this type of peaks for N$>$100, which actually correspond to
the close-packed (\emph{CP}) clusters changing to the \emph{DH}
motifs.

A few of structural types, $i.e$, polytetrahedra(\emph{PT}), polytetrahedral involve an ordered
array of disclinations(\emph{PT-d})
\emph{IH}, \emph{DH}, \emph{CP} and disordered(\emph{DIS})
structures, were found in present study. About these structures, the
detailed description can be found elsewhere,(e.g.,
Refs.\cite{Baletto05_review} and references therein) we briefly
summary here. \emph{PT} is made by packing five smallest tetrahedra
sharing a common edge. The \emph{PT} can further reduce its strain
by involve an ordered array of disclinations(hereafter labeled as
\emph{PT-d}), where there are more than five smallest tetrahedra
sharing a common edge. Both \emph{PT} and \emph{PT-d} can be
naturally divided up into tetrahedra with atoms at their corners.
 \emph{IH} can be decomposed into 20 fcc tetrahedra
sharing a common vertex in the central site. A close-shell \emph{IH}
has 20 triangular fcc(111) faces and 12 vertices. A decahedron is
made up of two pentagonal pyramids sharing a common basis. It has a
single fivefold axis and is formed by five tetrahedra sharing a
common edge along the fivefold axis. Both \emph{IH} and \emph{DH} can be have
close geometry shell structure. \emph{CP} is a pieces of fcc structure.
The truncated octahedron is one of \emph{CP}. By contrast, close-packed
structures are composed of octahedra and tetrahedra. \emph{DIS} is
that the clusters can hard be identified with any order structure.

Most \emph{PT} clusters are observed at small sizes. Specially all
clusters less than 24 atoms are \emph{PT} type, and most of them are
identical for all $\gamma$ at the same size. This means that the
cluster structures are less relevant with the details of potential.
Fig.3 shows the small size \emph{PT} clusters of N=4-23. These
clusters are on a \emph{PT} growth sequence. For example, the 7-atom
cluster has a structure in the shape of a pentagonal bipyramid,
which can be viewed as packing of five small tetrahedra by
introducing some inherent strains. The clusters larger than seven
atoms grow by introducing more new tetrahedra on its surface. The
\emph{PT} growth sequence keeps and leads to the 13-atom cluster,
which is commonly regarded as an icosahedron, however, it is also a
\emph{PT} according to Hoare's description, and can be view as the
packing of 20 small tetrahedra.\cite{Doye03_PT,Hoare71_PT} The
19-atom cluster is packed by small tetrahedra on the surface of the
13-atom cluster. The clusters grow further by packing around the
waist of the 19-atom cluster to larger clusters. The \emph{PT}
growth sequence is maintained and strain is also accumulated. There
are a few clusters in the size range smaller than 24, which did not
follow the growth sequence described above, but still has PT
structure. The exceptions are also listed in the Fig.3, which is
N=8,21,23 for $\gamma$=1.2,1.1, and 0.8 respectively. For cluster
with N=8 and $\gamma$=1.2, which is not the normal \emph{PT}
structure, is the fraction of a bcc structure. N=8 cluster can also
be viewed as a positive-disclination PT cluster(only four tetrahedra
sharing a common edge), it has more positive strain than normal PT.

Fig.4 presents the extension of the \emph{PT} growth sequence based
on the small size clusters. 34-atom cluster is complete form of the
waist-packing pattern based on the 19-atom cluster. The larger
\emph{PT} clusters($35<N<55$) are based on the 34-atom cluster,
which is similar to the growth mode form 13-atom \emph{PT} to
19-atom.

With increasing of cluster size, the positive strain is also
accumulated rapidly in \emph{PT}. To reduce the positive strain,
\emph{PT-d} was observed. Fig.5 presents a few selected \emph{PT-d}
clusters. This type of clusters has been discussed in several
papers.\cite{Doye97_morse,Doye01b_dzugutov,Doye03_PT} \emph{PT-d}
structure is different from \emph{PT} by introducing six tetrahedra
share a common edge in some sites. Although disclinations is unfavor
in local, it does result in the decrease of global strain of
clusters. In fact, this structure is similar to the square-triangle
Frank-Kasper phases usually used in the
quasi-crystals.\cite{Frank58_FK,Frank59_FK,Shoemaker88_FK,Shechtman84_quasi}

 The \emph{IH} clusters appear following the
\emph{PT-d} clusters at larger sizes. Some selected \emph{IH}
clusters are listed in the Fig.6. Each cluster is given in the side
view and the top view. There are many discussions in previous papers
about \emph{IH}.\cite{Baletto05_review} The \emph{IH} clusters can be
viewed as packing atoms based on a 13-atom pentadechedron but not a
13-atom Icosahedron. Usually, comparing with \emph{PT}, \emph{IH}
clusters have larger fcc(111) surface and shell-like structure.
These \emph{IH} clusters are stable due to their extreme large
fcc(111) surface and less strains comparing to \emph{PT}. For any
\emph{IH} based cluster, it can always form a closed-shell
icosahedron by packing enough atoms on it. There are also a few
\emph{IH} clusters(see Fig.7), which are interesting because they
are not the fracture of next closed-shell icosaheron, but the
fracture of the larger icosahedron. For example, \emph{IH} clusters
of N=48,50 are not the fracture of N=55 icosahedron, but the
fracture of N=147 icosahedron. Also clusters with
N=89,90($\gamma=0.8$) and N=62,65($\gamma=0.9-1.0$) are the fracture
of N=561 icosaheron. These clusters were discussed by Baletto
\emph{et al}\cite{Baletto01} as natural pathway to the growth of
larger \emph{IH} clusters.

Fig.8 shows some \emph{DH} ,\emph{CP} and \emph{DIS} structures.
Both \emph{DH} and \emph{CP} clusters are found for the small value
of $\gamma$. For simple pair potentials, the smaller \emph{DH} and
\emph{CP} can only exist for very narrow potential, or potential
with very large $d_{EPS}$, such as $C_{60}$ clusters. \emph{CP}
definitely have the same structure with fcc crystal, and \emph{DH}
is also more close to fcc structure over \emph{PT} and \emph{IH}.
Disorder clusters appear for larger $\gamma$ around 1.2, where the
potential is much flat around the bottom. Some disordered clusters
are in fact a serious distorted ordered structures. For example,
distorted \emph{IH}(N=85,103,$\gamma$=1.2) are listed. There are
many disordered structures are hard to be recognized based on any
ordered structures(N=38,$\gamma$=1.2). There are also some
interpenetrated clusters, which exhibit the combination of two type
structures.\cite{Doye97_morse} The  interpenetrated clusters were
found during the structure motif changing. The remarkable feature of
these clusters is that it exhibits different structural character by
looking from different direction, for examples, the cluster of N=59
and $\gamma=1.2$ is found between \emph{PT} and \emph{IH}.

Fig.9 plots the zero temperature structure 'phase diagram' as a
function of both cluster size and $\gamma$. It can be seen that, the
smaller $\gamma$, the more \emph{DH} and \emph{CP} clusters, and the
less disorder clusters. For $\gamma$=0.8, the structure is dominated
by the \emph{DH} and \emph{CP}. And only one \emph{DH} and one
\emph{CP} clusters are found in the potentials of $\gamma>0.9$. The
number of \emph{PT} clusters increase with increasing of $\gamma$.
For different potentials, this growth sequence stops at the
different sizes(N=24($\gamma$=0.8),N=31($\gamma$=0.9),
N=34($\gamma$=0.95),N=37($\gamma$=1.0,1.05,1.1),N=43($\gamma$=1.2)).
\emph{PT-d} clusters favor the size within 25 to 107. After the
\emph{PT-d} motif first emerges at $\gamma$=0.95, its number
increases with the increasing of $\gamma$, and reaches the maximum
around $\gamma$=1.05. As $\gamma$ continues to increase, the
distribution of \emph{PT-d} motif begins to reduce. At $\gamma$=1.2,
only seven \emph{PT-d} clusters are found. The disordered clusters
first emerge at the large size between the two magic number N=55 and
N=147 at $\gamma$=1.0, and its distribution increase as the
increasing of $\gamma$. At the value of $\gamma$=1.2, it is
dominated by the disordered cluster. The \emph{IH}-based clusters
usually appear in larger sizes. \emph{IH} clusters dominate for
value of $\gamma$=0.9 and 0.95, and its number decreases for both
$\gamma$ larger than 0.95 and smaller 0.9.

Since clusters have non-neglectful surface effect and large
deformation in contract with its bulk crystal, the competition
between the deformation(strain) and surface energy plays the key
role in determining cluster structures. According to this
consideration, we can give a qualitative explanation for the
existing of each structural type. Small size clusters have very
large surface-volume ratio, thus the surface effect is dominated.
\emph{PT} clusters are abundant at small sizes due to its the lowest
energy surface, fcc(111) surface. The dominant geometric effect can
not be easily affected unless for enough strong atomic interaction.
For Morse potential,\cite{Doye95_morse} only an extreme narrow
potential can breaks the \emph{PT} motif. Since strains in \emph{PT}
clusters increase rapidly due to continuously packing tetrahedra,
\emph{PT-d} replace the \emph{PT} with size increasing due to the
partially release of strains. Further increasing sizes, both
\emph{PT} and \emph{PT-d} are more and more unfavorable, thus the
\emph{IH} clusters appear. \emph{IH} clusters have the similar
surface as \emph{PT} and \emph{PT-d}, but the inner strain is much
reduced when taking \emph{IH} arrangement. This is the reason why
the \emph{IH} clusters appear after \emph{PT-d} ones.

Among \emph{IH}, \emph{DH} and \emph{CP} structure, \emph{IH} has
the lowest surface energy and largest deformation energy, \emph{CP}
is opposite, while \emph{DH} is in the middle. For small
$\gamma$(narrow potential), the structure with large
deformation(strain) is unfavorable, \emph{CP} and \emph{DH} are
dominated. The deformation(strain) can be neglected for largest
$\gamma$(more flat potential), the \emph{PT} and disordered clusters
are favorable. For middle value of $\gamma$, the \emph{PT-d} and
\emph{IH} are appeared. Generally \emph{DIS} should have the largest
strain energy, thus it can only survive for flat potential, this is
the case of large $\gamma$ in present case.  The cluster
distribution is in agreement with the Doye's qualitative principle
that decreasing the range of the pair-potential(the width of
potential well) has the effect of destabilizing strained structures.
It needs to point that a quantitative analysis is necessary for
detailed understanding.

The most stable structure of bulk phase is bcc for $\gamma>1$,
however only one cluster has the character as a fraction of bcc,
which is the one with N=8 and $\gamma$=1.2. On the contrary, a large
number of \emph{IH} and \emph{PT} clusters are found for $\gamma>1$.
The main reasons come from the surface effect of small cluster,
which result in the clusters adopting the most close packed fcc(111)
surface. However, there could be the inherent competition between
fcc and bcc, the existence of \emph{DIS} and \emph{PT-d} clusters
could be the results for $\gamma>1$.

\section{Summary}

In this paper, we have studied how the change of equilibrium pair
separation (or atomic size) for a fixed interaction range affects
the favored structures of atomic clusters. To do so, a simple pair
potential, which equilibrium pair separation can be varied under a
fixed interaction range by changing one of the potential parameters
has been proposed. This potential also can make both
face-centered-cubic structure and body-centered-cubic stable by changing
the parameter. The evolution of cluster structures are calculated
for several sets of parameters. Our results show that, the potential
with large equilibrium pair separation(larger atomic size) favors
\emph{DH} and \emph{CP} structures, disordered clusters appear for
the potential with small equilibrium pair separation, while for the
middle range of equilibrium pair separation, the \emph{IH}
structures are dominated. Present observation conforms the Doye and
Walse's qualitative principle that decreasing the range of the
pair-potential has the effect of destabilizing strained structures.

\begin{acknowledgments}
This research was supported by the National Science Foundation of
China, Shanghai Project for the Basic Research. The computation is
performed in the Supercomputer Center of Shanghai.

\end{acknowledgments}

\newpage

\begin{table}
  \centering
  \caption{The cohesive energy of both fcc and bcc for different
  sets of $\gamma$ and $\varepsilon$, where $E_{bcc}$ and $E_{fcc}$
  are the cohesive energy for bcc and fcc structures respectively,
  and $a_{bcc}$ and $b_{fcc}$ are the lattice constants for bcc and fcc structures
  respectively.
For $\gamma<1.0$, fcc phase is more stable, while $\gamma>1.0$ bcc
structure is more stable. At $\gamma=1.0$, fcc and bcc have the same
cohesive energy.}\label{tab1}
\begin{tabular}{|l|l|l|l|l|l|}
  \hline
  $\gamma$ & $\varepsilon$ &$E_{bcc}$&$E_{fcc}$&$a_{bcc}$& $a_{fcc}$\\\hline
        0.80&98.496 &-5.326&-6.000&2.222   &2.778\\\hline
        0.90&29.415 &-5.762&-6.000&2.036   &2.557\\\hline
        0.95&18.760 &-5.848&-6.001&1.932   &2.457\\\hline
        1.00&12.786 &-6.036&-6.036&1.859   &2.361\\\hline
        1.05& 9.168 &-6.178&-6.124&1.793   &2.264\\\hline
        1.10& 6.844 &-6.292&-6.252&1.730   &2.176\\\hline
        1.20& 4.172 &-6.604&-6.547&1.589   &2.021\\\hline
\end{tabular}
\end{table}
\newpage

\begin{figure}[fig1]
\centering
\includegraphics[angle=-90,width=100mm]{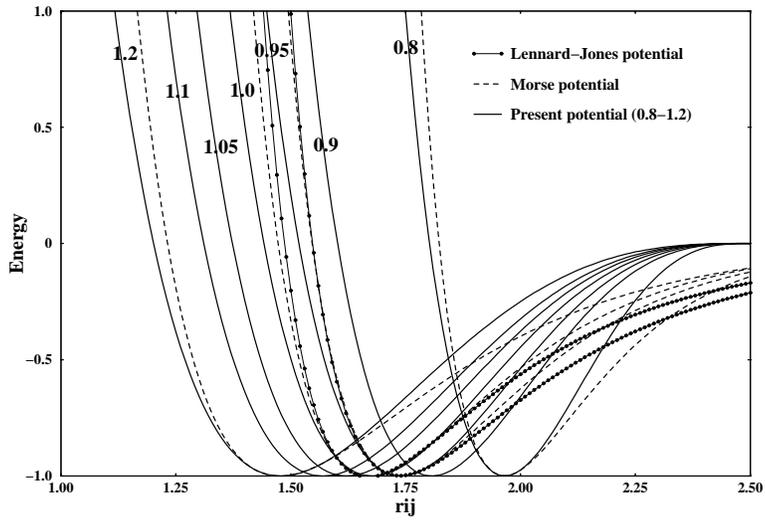}
\caption{The new potential(solid line) with a few selected $\gamma$
of 0.8,0.9,0.95,1.0,1.05,1.1,1,2. For comparing, Morse(dashed line)
and LJ(dotted-solid line) potential are also shown, which has been
fitted to have the same curvature at the bottom of the potential
well as the present one.}
\end{figure}

\begin{figure}[fig2]
\centering
\includegraphics[angle=-90,width=100mm]{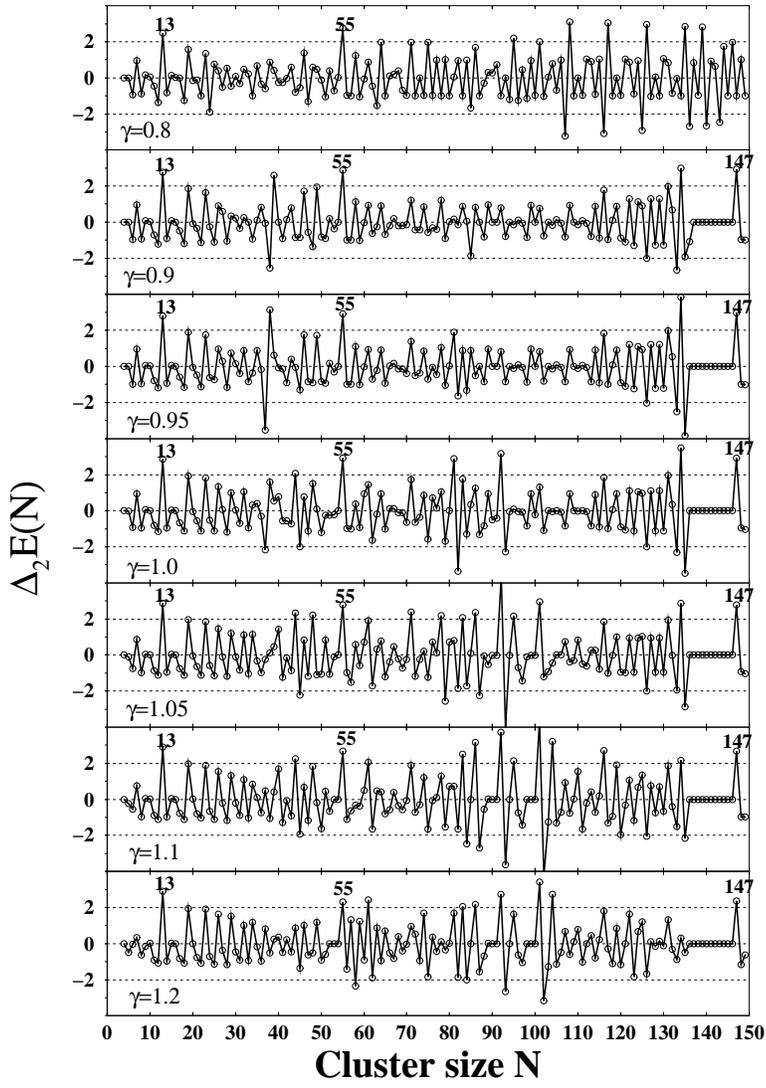}
\caption{The second energy difference ($\Delta_{2}E$) as a function
of cluster size for all $\gamma$. Peaks in $\Delta_{2}E$ correspond
to clusters which are stable compared to adjacent sizes.}
\end{figure}

\begin{figure}
\centering
\includegraphics[width=150mm]{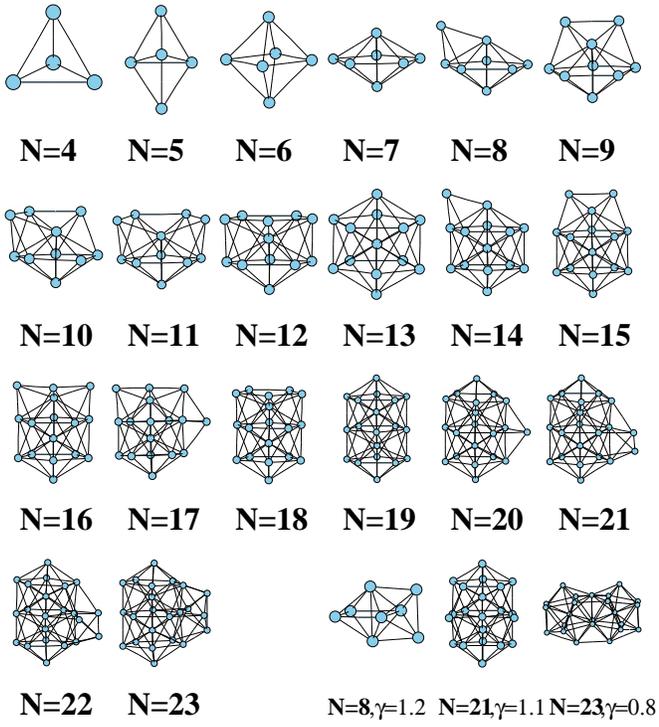}
\caption{The small size polytetrahedral clusters of N=4-23. Most of
them are identical for all $\gamma$ at the same size. The exceptions
are N=8,21,23 for $\gamma$=1.2,1.1, and 0.8 respectively. }
\end{figure}

\begin{figure}[fig4]
\centering
\includegraphics[width=150mm]{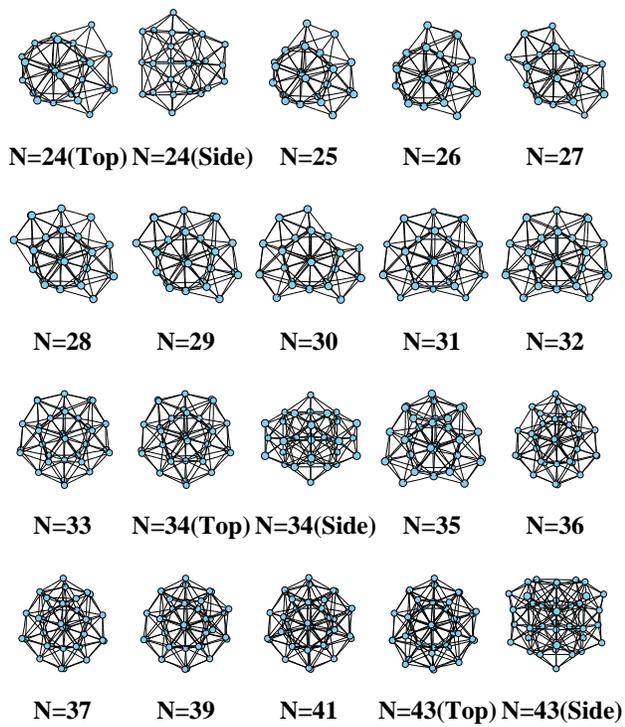}
\caption{The extension of the polytetrahedral growth sequence based
on the small size clusters.}
\end{figure}

\begin{figure}[fig5]
\centering
\includegraphics[width=150mm]{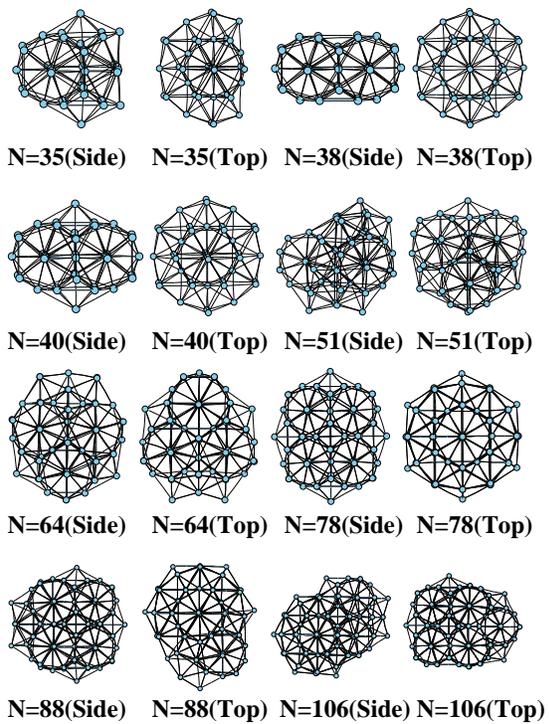}
\caption{A few selected \emph{PT-d} clusters.}
\end{figure}

\begin{figure}[fig6]
\centering
\includegraphics[width=150mm]{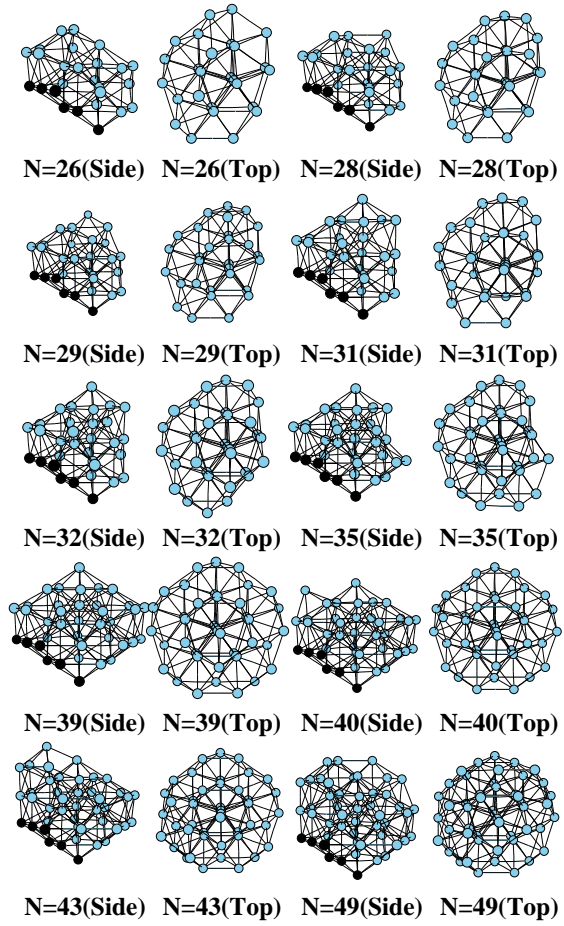}
\caption{Some selected \emph{IH} clusters. Each cluster presented in the
side view and the top view, with a 6-atom fcc(111) facet marked in the
side view.}
\end{figure}

\begin{figure}[fig7]
\centering
\includegraphics[width=150mm]{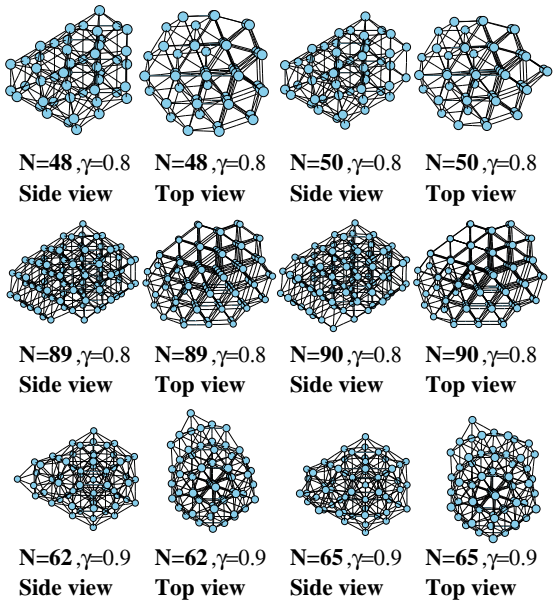}
\caption{Some \emph{IH} clusters are a fraction of larger
closed-shell icosahedrons. The clusters with N=48 and 50 are the
fracture of N=147 icosahedron; and N=89, 90($\gamma=0.8$) and
N=62,65($\gamma=0.9-1.0$) are the fracture of N=561 icosahedron. }
\end{figure}

\begin{figure}[fig8]
\centering
\includegraphics[width=150mm]{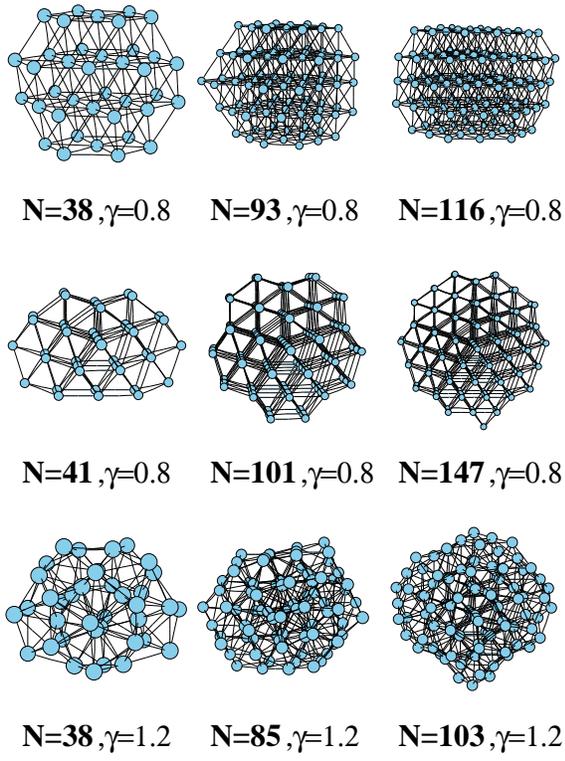}
\caption{Some \emph{DH}, \emph{CP} and \emph{DIS} clusters,
Top row: \emph{CP} clusters; Middle row: \emph{DH} clusters;
Bottom row: \emph{DIS} clusters.}
\end{figure}

\begin{figure}[fig9]
\centering
\includegraphics[angle=-90,width=150mm]{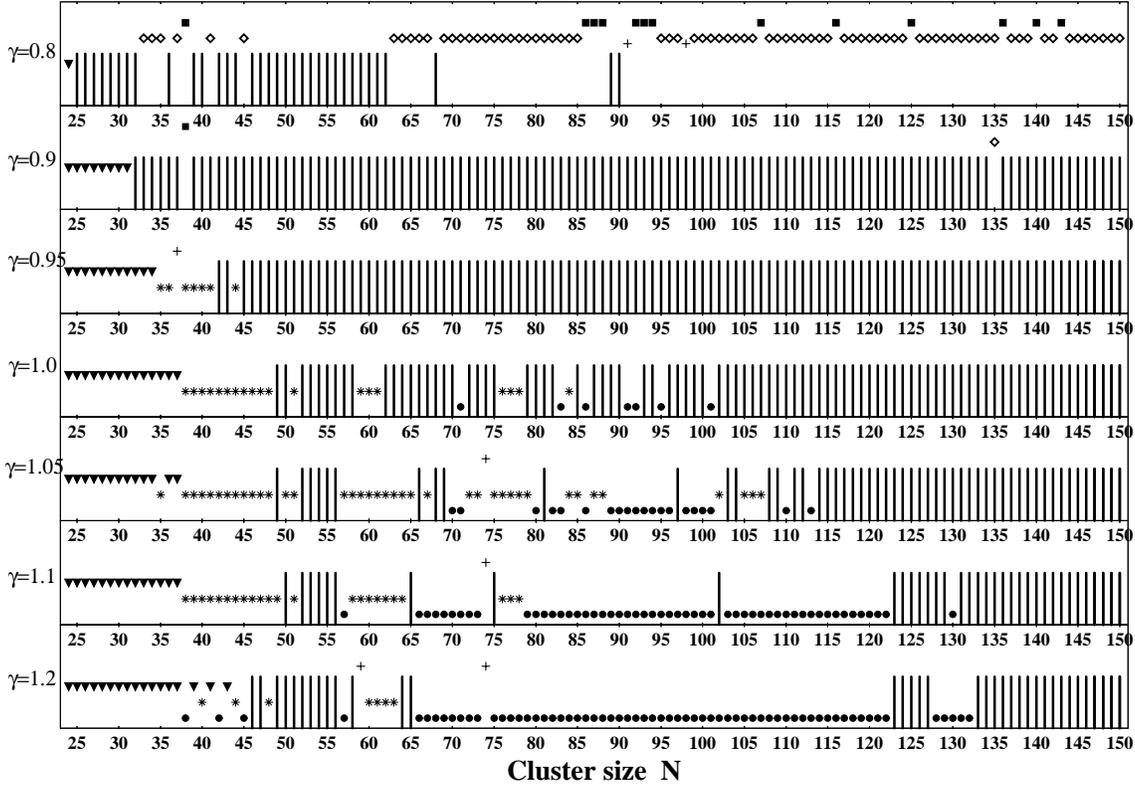}
\caption{The zero temperature structural 'phase diagram' as a
function of both size and $\gamma$.
From left, 'Filled triangles' for \emph{PT},
'Stars' for \emph{PT-d} and 'Vertical lines' for \emph{IH}.
'Filled square' for \emph{CP}, 'Open diamond' \emph{DH},
'Plus' for some minor structure including interpenetrated clusters,
a truncate tetrahedra(N=91, $\gamma$=0.8),\cite{Doye01_c60} and a rare
tetrahedron(N=98, $\gamma$=0.8).\cite{Leary99_lj}}
\end{figure}

\end{document}